\documentstyle[psfig,amssymb,referee]{mn}

\title[Spectroscopic constraints on the stellar population of ellipticals]
{ Spectroscopic Constraints on the Stellar Population of Elliptical Galaxies
in the Coma Cluster}

\author[B. Mobasher \& P. A. James] 
{B. Mobasher$^{1,2}$ and P. A. James$^3$\\  
$^1$Astrophysics Group, Blackett Laboratory, Imperial College, 
Prince Consort Rd, London SW7 2BZ, UK\\
$^2$ Space Telescope Science Institute, 3700 San Martin Drive, Baltimore, 
MD 21218, USA\\  
$^3$ Astrophysics Research Institute, Liverpool 
John Moores University, Egerton Wharf, Birkenhead, CH41 1LD, UK\\
}
\begin{document} 
\maketitle
\begin{abstract} 

Near-IR spectra for a sample of 31 elliptical galaxies in the Coma
cluster are obtained. The galaxies are selected to be ellipticals (no
lenticulars), with a large spatial distribution, covering both the
core and outskirt of the cluster (ie. corresponding to regions with
large density contrasts). CO$_{sp}$ (2.3 $\mu$m)
absorption indices], measuring the
contribution from intermediate-age red giant and supergiant stars to
the near-IR light of the ellipticals, are then estimated.

It is found that the strength of CO$_{sp}$ features in elliptical
galaxies increases from the core ($r < 0.2^\circ$) to the outskirts 
($r > 0.2^\circ$)
of the Coma cluster. Using the Mg$_2$ strengths, it is shown that the
observed effect is not due to metallicity and is mostly caused by the
presence of a younger population (giant and supergiant stars) in
ellipticals in outskirts (low density region) of the cluster.

Using the CO$_{sp}$ features, the origin of the scatter on the near-IR
Fundamental Plane (relation between the effective diameter, effective surface
brightness and velocity dispersion) 
of elliptical galaxies is studied. Correcting this relation
for contributions from the red giant and supergiant stars, the {\it rms}
scatter reduces from 0.077dex to 0.073dex. Although measurable, 
the contribution from these intermediate-age stars to the scatter on the 
near-IR Fundamental Plane of ellipticals is only marginal. 

A relation is found between the CO$_{sp}$ and $V-K$ colours of
ellipticals, corresponding to a slope of $0.036\pm 0.016$,
significantly shallower than that from the Mg$_2-(V-K)$ relation. This
is studied using stellar synthesis models.

\end{abstract}

\begin{keywords}
  galaxies: clusters - galaxies: elliptical - galaxies: fundamental parameters
  - galaxies: stellar content- infrared: galaxies   
\end{keywords}

\section {Introduction}

There is growing evidence that early-type galaxies located in small groups
and in the field may possess an intermediate-age stellar component
($\sim 1-5$ Gyrs), in addition to the old stellar population 
characteristic of their cluster counterparts  
(Bothun \& Gregg 1990; Caldwell et al. 1996).  For example,  
using the strength of optical spectral lines, the presence of a substantial 
population of intermediate-age stars is confirmed in elliptical galaxies
in low density environments, an effect which is much reduced in 
ellipticals in denser regions (Bower et al 1990; Rose et al 1994). 
This is further supported by spectroscopic ($H\beta$) and photometric
(UV-optical colours) observations of field ellipticals, showing evidence for
star formation in these galaxies within the last $3-5$ Gyrs
(Schweizer \& Seitzer 1992; Gonzalez 1993).  

This systematic 
difference in age poses serious problems in using relations established 
for cluster ellipticals to study field galaxies. For example, the
Fundamental Plane (FP) relation between the velocity dispersion, 
effective surface brightness and effective radius of ellipticals, used
to measure the peculiar velocity field and to study the evolution
and formation of galaxies, is often zero-pointed using cluster samples. 
Any difference between the cluster and field populations will lead 
to spurious results. The effect of this intermediate-age stellar
population on the FP of ellipticals in low density environments is 
investigated by Guzm\'an et al (1994). 

Recently, an attempt has been made to reduce the contribution from
this young population on the FP of ellipticals by extending this
relation to near-IR ($2.2\ \mu m$) wavelengths (Pahre et al 1998; Mobasher
et al 1999). The near-IR FP is expected to be less
affected by differences in age, metallicity and
stellar population among the ellipticals, compared to its optical
counterpart. Therefore, the relatively large {\it rms} scatter found
for the near-IR relation (0.076 dex and 0.074 dex for the
infrared and optical relations respectively) is surprising. This is
either due to contributions to their near-IR light from the red giant
and supergiant stars, or is caused by differences in matter
distribution and the internal dynamics (i.e. orbital anisotropy or
rotation) among the ellipticals. Understanding the origin of this
scatter is essential in constraining models of formation and evolution
of elliptical galaxies.

A good indicator of the presence of an intermediate-age stellar
population in elliptical galaxies is the strength of their
spectroscopic near-IR CO (2.3 $\mu m$) absorption feature, since this
is mainly produced in the atmosphere of giant and supergiant stars.
Also, unlike $Mg_2$ line indices, these are relatively insensitive to
the on-going star formation (ie. formation of main sequence stars
prior to the formation of the giant and supergiant populations) 
and extinction by dust.  In a
recent study, the value of near-IR CO bandheads in constraining
stellar population of elliptical galaxies is explored (Mobasher \&
James 1997; James \& Mobasher 1999). Measuring CO strengths for a
small and heterogeneous sample of cluster and field ellipticals, the
presence of a younger component in at least some field ellipticals was
confirmed. Here, we extend this study to ellipticals 
at the core and outskirts of the Coma cluster. 
This provides a homogeneous sample for studying the
stellar population of ellipticals in regions with large density
contrasts. By choosing the ellipticals at the core and outskirt of a
single cluster, we aim to minimise the contribution due to 
chemical evolution in galaxies as a function of environment, with the
{\it only} difference being the local density.

Section 2 presents the spectroscopic observations and data reduction.
A discussion of the physical significance of the strength of near-IR
CO features is given in section 3. This is followed in
section 4 by a study of the radial dependence of CO strengths in the
Coma ellipticals. Section 5 explores the source of the scatter in the
near-IR FP of ellipticals. The relation between the strength of CO
features and other photometric parameters in ellipticals is studied in
section 6. The conclusions are summarised in section 7.

\section {Sample Selection, Observations and Data Reduction}

The galaxies for this study are selected to be 
ellipticals (i.e. no lenticulars), confirmed members of the Coma
cluster and have a wide enough spatial distribution to cover both the
core and outskirts of the cluster (i.e. regions with large density
contrast). Other spectroscopic (velocity dispersion, $Mg_2$ line
strengths) and photometric (optical and near-IR luminosities) data are
available for all the galaxies in the sample.

The observations were carried out using the United Kingdom Infrared
Telescope (UKIRT) during the 4 nights of 21--24 February 1999. The
instrument used was the long-slit near-IR spectrometer CGS4, with the
40~line~mm$^{-1}$ grating and the long-focal-length (300~mm) camera.
The 4-pixel-wide slit was chosen, corresponding to a projected width
on the sky of 2.4~arcsec. Working in 1st order at a central wavelength
of 2.2 $\mu m$, this gave coverage of the entire K window. The CO
absorption feature, required for this study, extends from 2.293 $\mu
m$ (rest frame) into the K-band atmospheric cut-off. The principal
uncertainty in determining the absorption depth comes
from estimating the level and slope of the continuum shortward of this
absorption which requires wavelength coverage down to at least 2.2
$\mu m$ and preferably shorter. There are many regions of the
continuum free from lines even at this relatively low resolution. The
effective resolution, including the degradation caused by the wide slit, 
is about 230.

For each observation, the galaxy was centred on the slit by maximising
the IR signal, using an automatic peak-up facility.  Total on-chip
integration times of 12 minutes were used for the
brightest and most centrally concentrated ellipticals while an
integration time of 24 minutes was more typically required.
During this time, the galaxy was slid up and down the slit at one
minute intervals by 22~arcsec, giving two offset spectra which were
subtracted to remove most of the sky emission. Moreover, the array was
moved by 1 pixel between integrations to enable bad pixel replacement
in the final spectra. Stars of spectral types A0--A6,
suitable for monitoring telluric absorption, were observed in the same
way before and after each galaxy, with airmasses matching those of the
galaxy observations as closely as possible. Flat fields and argon arc
spectra were taken using the CGS4 calibration lamps. A total of 31
elliptical galaxies in the core and outskirts of the Coma cluster were
observed.

The data reduction was performed using the FIGARO package in the
STARLINK environment. The spectra were flatfielded and a polynomial
was fitted to estimate and remove the sky background. These spectra
were then shifted to the rest frame of the galaxy, using their
redshifts. The atmospheric transmissions 
were corrected by dividing the spectra with
the spectrum of the standard star observed
closely in time to the galaxy, and at a similar airmass. The resulting
spectra was converted into a normalised, rectified spectrum by fitting
a power-law to featureless sections of the continuum and dividing the
whole spectrum by this power-law, extrapolated over the full
wavelength range. 

To measure the depth of the CO absorption feature, the same procedure
outlined in James and Mobasher (1999) is used.  The restframe,
rectified spectra were rebinned to a common wavelength range and
number of pixels, to avoid rounding errors in the effective wavelength
range sampled by a given number of pixels.  Two methods were then used
to define the CO strength for each spectrum.  The first is the
spectroscopic CO index (Doyon et al. 1994), CO$_{sp}$, which is the
mean level of the rectified spectrum, between wavelength limits of
2.31~$\mu$m and 2.4~$\mu$m, expressed as a magnitude difference,
relative to the continuum level.  The second measurement was the CO
equivalent width (Puxley, Doyon \& Ward 1998), CO$_{EW}$, 
which quantifies
the depth of the CO absorption between 2.293~$\mu$m and 2.32~$\mu$m.
Both CO$_{sp}$ and CO$_{EW}$ are defined such that a deeper absorption
corresponds to a larger number.  CO$_{sp}$ has the advantage that the
fractional Poisson errors are decreased by averaging the absorption
over a larger wavelength range, whereas CO$_{EW}$ is claimed to be more
sensitive to stellar population variations, and is less subject to
errors in the power-law fitting.  Also, CO$_{EW}$ can be used for higher
redshift galaxies, due to the shorter wavelength range, although that
is not a consideration for the present study. 
 
There are three principal and quantifiable sources of error in the measured
CO$_{sp}$ values here. The first is due to pixel-to-pixel noise in the
reduced spectra, as calculated from the standard deviation in the
fitted continuum points, assuming that the noise level
remains constant through the CO absorption. This gives an error on both
the continuum level and the mean level in the CO$_{sp}$ absorption, which
were added in quadrature.  The second error component comes from the
formal error provided by the continuum fitting procedure.  This
could leave a residual tilt or curvature in the spectrum. 
The formal error was used to quantify this contribution.  The
final component is an estimate of the error induced by redshift and
wavelength calibration uncertainties. All three errors were of similar
sizes, and when added in quadrature give a typical uncertainty in
CO$_{sp}$ of $\pm 0.012$ mag.
Furthermore, we repeated one galaxy (NGC 2832) on separate nights. The
completely independent reductions of the two observations 
gave CO$_{sp}$ indices corresponding to  0.257 mag. and 0.263 mag. 
This is consistent
with the random components of the error calculation above, and well
within our total error estimate.

The strength of CO absorption features and their corresponding
equivalent widths for the Coma ellipticals observed in this study are
presented in Table 1. Column 2 lists the radial distance (in degrees)
from the core of the Coma cluster. Columns 3-6 give, respectively, 
the total near-IR magnitudes ($K_{tot}$), velocity dispersions
($log(\sigma)$), optical-IR colours ($V-K$) and spectroscopic Mg$_2$
measurements, all taken directly from Table 1 in Mobasher et al (1999). 
Finally, estimates of the CO$_{sp}$ and CO$_{EW}$ are given in 
columns 7 and 8 respectively. 

To allow the comparison between the CO$_{sp}$ features from the
present sample and those for higher redshift ellipticals
(for which only CO$_{EW}$ is measurable), the relation
between the CO strength of absorption feature and the CO equivalent
width for individual galaxies is presented in Figure
1. A least squares fit to this relation gives

$$   CO_{sp} = (0.210\pm 0.005)\ log\ (CO_{EW}) + (0.136\pm 0.005) $$
with {\it rms =0.008}. The deviant point in Figure 1 is N4971, which is
excluded from the fit.  For the analysis in the following sections, the
CO$_{sp}$ strengths are used (in magnitude units). For nearby
galaxies, where the CO$_{sp}$ is not heavily contaminated by
atmospheric lines, this is a reasonable procedure. The 
subsequent results in this paper do
not depend on whether the CO$_{sp}$ or CO$_{EW}$ are used. 

\begin{figure} 
\centerline{\psfig{figure=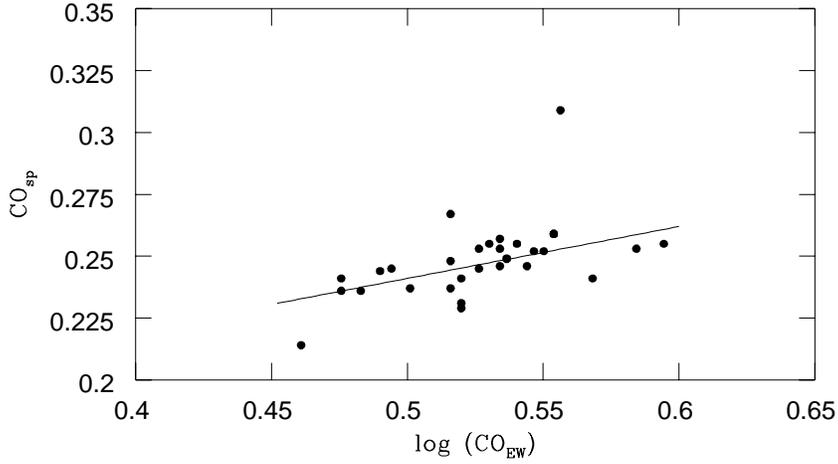,width=0.9\textwidth,angle=0}}
\caption{The CO$_{sp}-$CO$_{EW}$ relation for the sample of elliptical
galaxies in Table 1. The linear least squares fit to the data is also
shown. The galaxy N4971 is $> 3\sigma$ deviant from the mean relation
and is not included in the fit. 
}
\end{figure}

\begin{table}
\caption{The photometric and spectroscopic parameters
for elliptical galaxies in the Coma cluster}
\begin{tabular}{lcccccccc}\\ 
 & & & & & & &\\
Name      & r &  $K_{tot}$& $log(\sigma)$ & $V-K$ & Mg$_2$ & CO$_{EW}$ &
CO$_{sp}$ \\  
& & & & & & &\\
 RB45      &   0.040 &  11.91 &  2.133 & 3.10 &   0.280 & 3.31 &  0.231\\
 N4886     &   0.050 &  10.86 &  2.209 & 2.95 &   0.254 & 3.70 &  0.241\\
 RB43      &   0.056 &  12.19 &  2.230 & 3.01 &   0.262 & 2.99 &  0.236\\
 N4889     &   0.060 &   8.20 &  2.595 & 3.32 &   0.348 & 3.55 &  0.252\\
 N4874     &   0.060 &   8.55 &  2.398 & 3.26 &   0.323 & 3.31 &  0.229\\
 N4876     &   0.062 &  10.89 &  2.267 & 3.19 &   0.242 & 2.89 &  0.214\\
 IC4011    &   0.064 &  11.78 &  2.061 & 3.04 &   0.279 & 3.17 &  0.237\\
 N4872     &   0.069 &  11.30 &  2.329 & 3.09 &   0.300 & 3.50 &  0.246\\
 IC4021    &   0.112 &  11.58 &  2.205 & 3.23 &   0.299 & 3.09 &  0.244\\
 N4869     &   0.119 &  10.27 &  2.303 & 3.23 &   0.315 & 3.93 &  0.255\\
 IC4012    &   0.125 &  11.40 &  2.258 & 3.30 &   0.292 & 3.42 &  0.253\\
 N4867     &   0.135 &  11.12 &  2.352 & 3.17 &   0.307 & 2.99 &  0.241\\
 N4864     &   0.144 &  10.15 &  2.289 & $-$  &   0.286 & 3.42 &  0.257\\
 N4906     &   0.181 &  10.89 &  2.228 & 3.17 &   0.288 & 3.42 &  0.246\\
 D204      &   0.404 &  11.84 &  2.114 & 3.05 &   0.268 & 3.04 &  0.236\\
 D160-100  &   0.441 &  11.47 &  2.269 & 3.20 &   0.285 & 3.28 &  0.267\\
 N4926     &   0.565 &   9.80 &  2.420 & 3.32 &   0.324 & 3.47 &  0.255\\
 N4841A    &   0.723 &   9.53 &  2.414 & 3.24 &   0.320 & 3.58 &  0.259\\
 D140      &   0.745 &  11.42 &  2.232 & 3.13 &   0.297 & 3.84 &  0.253\\
 D160-27   &   0.768 &  11.30 &  2.235 & 3.14 &   0.282 & 3.31 &  0.241\\
 N4816     &   0.839 &   9.95 &  2.330 & 3.24 &   0.310 & 3.58 &  0.259\\
 IC4133    &   0.879 &  11.18 &  2.233 & 3.14 &   0.289 & 3.44 &  0.249\\
 N4807     &  1.067 &  10.28 &  2.310 & 3.17 &   0.285 & 3.36 &  0.245\\
 IC843     &  1.221 &  10.02 &  2.389 & 3.51 &   0.303 & 3.52 &  0.252\\
 N4789     &  1.525 &   9.30 &  2.416 & 3.25 &   0.304 & 3.44 &  0.249\\
 N4971     &  1.656 &  10.57 &  2.250 & 3.28 &   0.291 & 3.60 &  0.309\\
 D159-83   &  2.500 &  10.23 &  2.306 & 3.38 &   0.275 & 3.39 &  0.255\\
 D160-159  &  2.790 &  10.66 &  2.358 & 3.24 &   0.280 & 3.28 &  0.237\\
 N4673     &  3.294 &   9.68 &  2.345 & 3.18 &   0.270 & 3.36 &  0.253\\
 D159-43   &  4.545 &  10.60 &  2.399 & 3.33 &   0.338 & 3.28 &  0.248\\
 D159-41   &  4.750 &  11.11 &  2.277 & 3.23 &   0.324 & 3.12 &  0.245\\

\end{tabular}
\end{table}

\vfil\break

\section {What do Near-IR CO indices measure in 
elliptical galaxies ?}

The CO band at $2.3 \mu m$, lies longward of 2.29 $\mu m$
and constitutes the strongest
absorption feature in the K spectrum.
Beyond about 2.5~$\mu$m, the near-infrared spectrum is
contaminated by the atmospheric OH lines, producing spurious
absorption features and low atmospheric
transmission. The 2.3 $\mu m$ CO absorption feature is present in the
atmosphere of red giant (including Asymptotic Giant Branch- AGB) and
supergiant stars. Its depth increases with decreasing stellar
temperature, increasing stellar luminosity (Kleinman \& Hall 1986),
and increasing metallicity (Aaronson et al 1978). This implies that
red giant and supergiant stars have deeper CO absorption features than
dwarf stars. Moreover, supergiants are expected to have stronger CO
features than giant stars of the same temperature (this is because the
former have a higher microturbulent velocity, implying that the CO
absorption band is made of many saturated lines, leading to reduced
dependence of metallicity on the CO strength). For example, Doyon et
al (1994) found that the strength of the CO band associated with a
young stellar population reaches a maximum between 15 and 40 Myrs and
a CO$_{sp}$ index of $\sim 0.28$ mag. This is $\sim 0.1$ mag higher
than that observed for normal galaxies and is due to a
contribution from red supergiants.

Since the CO strengths provide a diagnostic for identifying the 
young to intermediate age AGB and supergiant population in galaxies, and
because of its relative insensitivity to non-stellar radiation and
dust reddening, these features have been widely used to identify
stellar populations in dusty infrared luminous galaxies (i.e. starbursts)- 
(Goldader et al 1997; Doyon et al 1994; Ridgway et al 1994). 
However, such studies are
complicated by the fact that the CO strength also depends to some
extent on the metallicity, in spite of the effects noted in the
previous paragraph.   
This effect has been 
studied by Frogel, Cohen \& Persson (1983), using globular clusters
with measured metallicities and photometric CO indices. Using this
calibration and the transformation between photometric and 
spectroscopic CO features, Doyon et al (1994) find 
$\Delta (CO_{sp}) = 0.11 \Delta [Fe/H], $ where $[Fe/H]$ is the logarithm
of metal abundance relative to the Sun. 

To use the CO strengths
to study evolutionary properties of elliptical galaxies, 
it is therefore important to separate the relative 
contributions from metallicity and stellar population. There is currently
no population synthesis models for composite stellar systems 
with satisfactory treatment of the intermediate age 
AGB and red supergiants and hence, near-IR CO features. 
Using the sample of Coma ellipticals in
Table 1, we find a relation between their CO$_{sp}$ and Mg$_2$ 
strengths (Figure 2). 
A linear fit to the 30 galaxies (excluding N4971) 
with available CO$_{sp}$ and Mg$_2$ gives
$$CO_{sp} = (0.189\pm 0.073) Mg_2 + (0.191 \pm 0.003) $$
with {\it rms} = 0.01. 
Assuming Mg$_2$ to be a measure of metallicity, the trend in Figure 2
demonstrates changes in CO indices due to metallicity, while the scatter
in this diagram, at a given Mg$_2$, indicates variations in CO indices
due to contributions from AGB and supergiant stars to the near-IR light
of ellipticals (this also includes measurement errors in CO indices).  
The slope here is steeper than 0.11 
found by Doyon et al. (1994)
which was estimated from the globular clusters, with most of them having
sub-solar metallicities, requiring extrapolation
beyond solar metallicity. 
Moreover, the Mg$_2$ indices
for elliptical galaxies are also   
likely to be affected by the young population
(i.e. residual star formation). Nevertheless, we use the relation in
Figure 2 
in subsequent sections to explore changes in CO$_{sp}$ strengths 
due to age and metallicity and to separate these effects from that due to 
contributions from the giant and supergiant stars.  

\begin{figure} 
\centerline{\psfig{figure=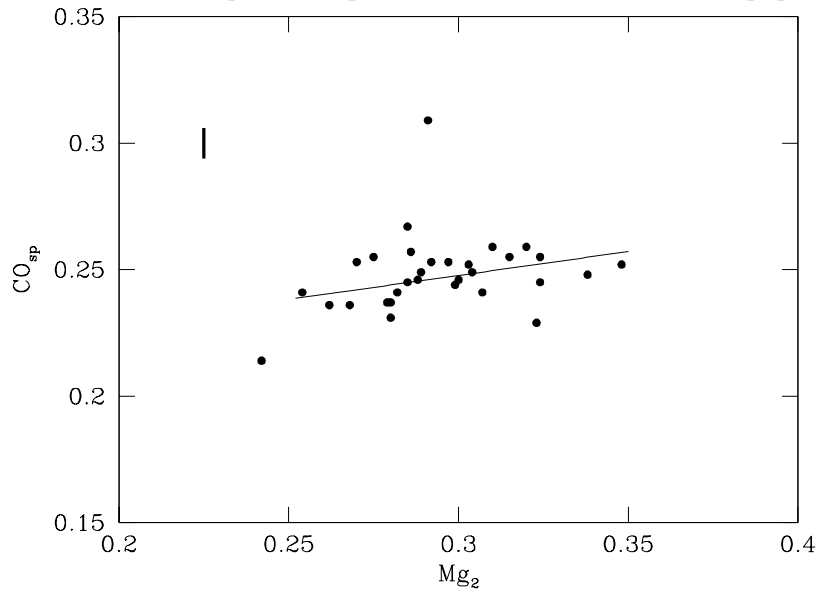,width=0.9\textwidth,angle=0}}
\caption{The CO$_{sp}-$Mg$_2$ relation for elliptical galaxies 
in the Coma cluster (Table 1). The linear least squares fit, 
excluding the deviant point (N4971), is also shown. The errorbar shows the
uncertainty in CO$_{sp}$ measurement for individual galaxies. }
\end{figure}

\section { Spectroscopic evolution of ellipticals as a function of 
environment}

The homogeneity of the present sample, combined with its
spatial coverage, allows a study of the spectroscopic properties of
ellipticals with radial distance (i.e. local density) from the core of
the Coma cluster. 
The CO$_{sp}$ histograms for elliptical galaxies at the
core ($r < 0.2^\circ$) and outskirt ($r > 0.2^\circ$) of the Coma
cluster are compared in Figure 3a, with their mean $<CO_{sp}>$ values
listed in Table 2. This shows that, on average, 
the ellipticals in the outskirts of
the Coma cluster have a stronger CO$_{sp}$ feature compared to those at
the core. This is a $\sim 3\sigma$ effect, with $< 1\%$ chance of
them belonging to the same parent population. 
Furthermore, considering a standard deviation of 0.012 mag. 
in the estimated CO$_{sp}$ values for individual galaxies, 
the observational errors corresponding to mean CO$_{sp}$ at the core 
(17 galaxies) and outskirt (14 galaxies) of the Coma cluster correspond 
to 0.0029 mag. and 0.0032 mag. respectively. These estimates are 
fully consistent
with the errors quoted in Table 2. However, they imply that the bulk of the
scatter is due to measurement errors.

For the same galaxies,
distribution of the Mg$_2$ indices are also compared in Figure 3b with
the mean $<Mg_2>$ values listed in Table 2. There are no significant
changes in $<Mg_2>$ strengths with radial distance from the core of
the Coma. This implies that the observed difference in CO$_{sp}$
strengths is not likely to be due to metallicity and is mainly caused by
changes in contributions from the intermediate age AGB and
supergiant population among the ellipticals at different distances
from the cluster core. 

\begin{table}
\caption{ Mean CO$_{sp}$ and Mg$_2$ estimates
for ellipticals at the core and outskirt
 of the Coma cluster and a 
sample of $z=0.07$ clusters.
}
\begin{tabular}{lccc}\\ 
 & & & \\
               &   $r < 0.2^\circ$  &  $r > 0.2^\circ$  &     Total\\
Coma           &                    &                   &          \\      
               &                    &                   &          \\
$< CO_{sp} >$  &   $0.241\pm 0.003$ &  $ 0.254\pm 0.004$ &   $0.248\pm 0.005$\\
               &                    &                    &      \\
$< Mg_2 >$     &   $0.291\pm 0.007$ &  $0.297\pm 0.005$  &   $0.294\pm 0.009$\\
               &                    &                    &       \\

Pisces/A2199/A2634 &  &  &\\
               &                    &                   &          \\
$< CO_{sp} >$  &                    &                   &  $0.244\pm 0.004$\\
\end{tabular}
\end{table}

The conclusion is that the ellipticals at low
density regions of the Coma (ie. cluster outskirts) have a population of AGB
and/or supergiant stars, not present in ellipticals at the denser
(core) region, indicating that the galaxies in the outer parts of the
Coma cluster are relatively younger than those at the core.  

\begin{figure} 
\centerline{\psfig{figure=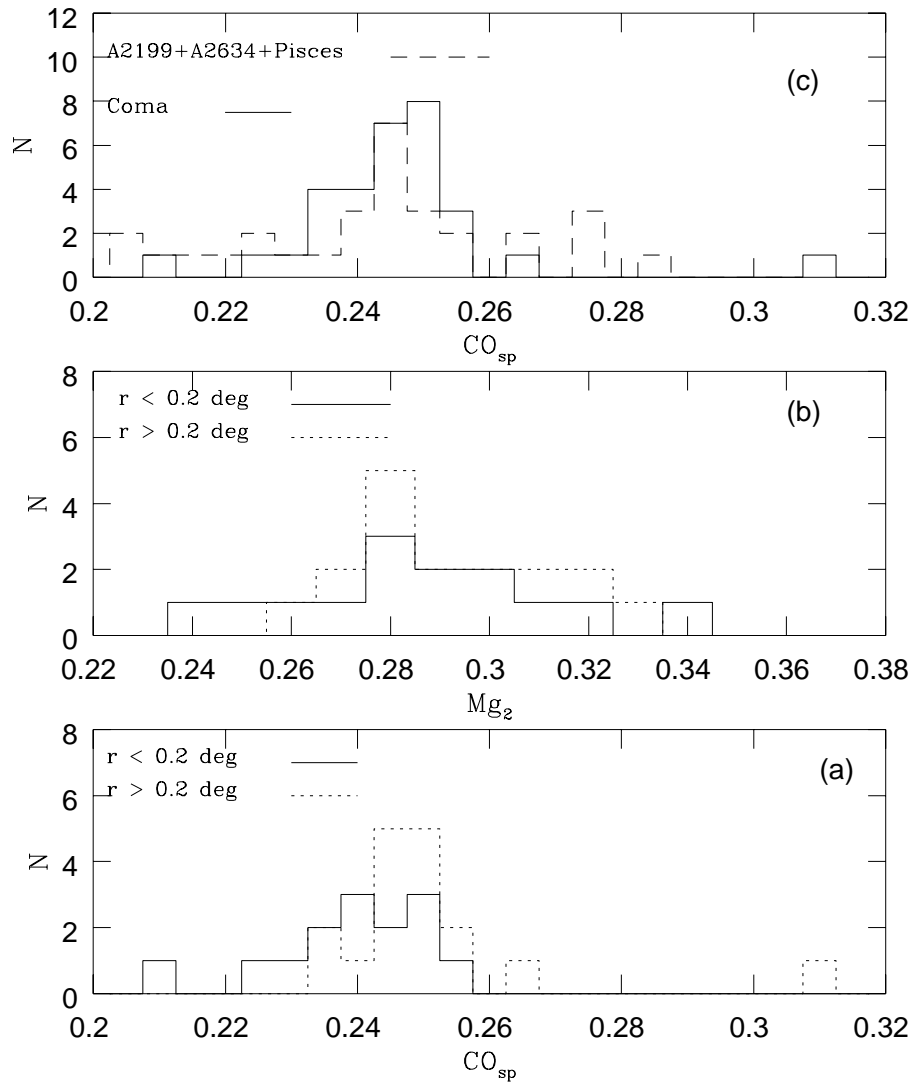,width=1\textwidth,angle=0}}
\caption{(a). Distribution of CO$_{sp}$ for ellipticals at the core
(solid line) and outskirts (dotted line) of the Coma cluster.
(b). The same as 3a for Mg$_2$ indices. (c). Comparison between the
CO$_{sp}$ distribution for elliptical galaxies in the Coma
(both core and outskirts)-(solid line) and the sample of ellipticals
in A2199, A2634 and Pisces clusters at $z\sim 0.07$, taken from 
James \& Mobasher (1999)-(dashed line).}  
\end{figure}

The variation in the CO$_{sp}$ with the radial 
distance from the cluster core, shown in figure 4
The average scatter in the CO$_{sp}$, at a given radius, is 0.004 mag
(corresponding to the scatter in the zero-point in the above
relation), which is 
significantly less than the variation in CO$_{sp}$ strength across the
cluster, listed in Table 2. This confirms again the radial
dependence of the CO$_{sp}$ features for ellipticals in the Coma cluster. 

\begin{figure} 
\centerline{\psfig{figure=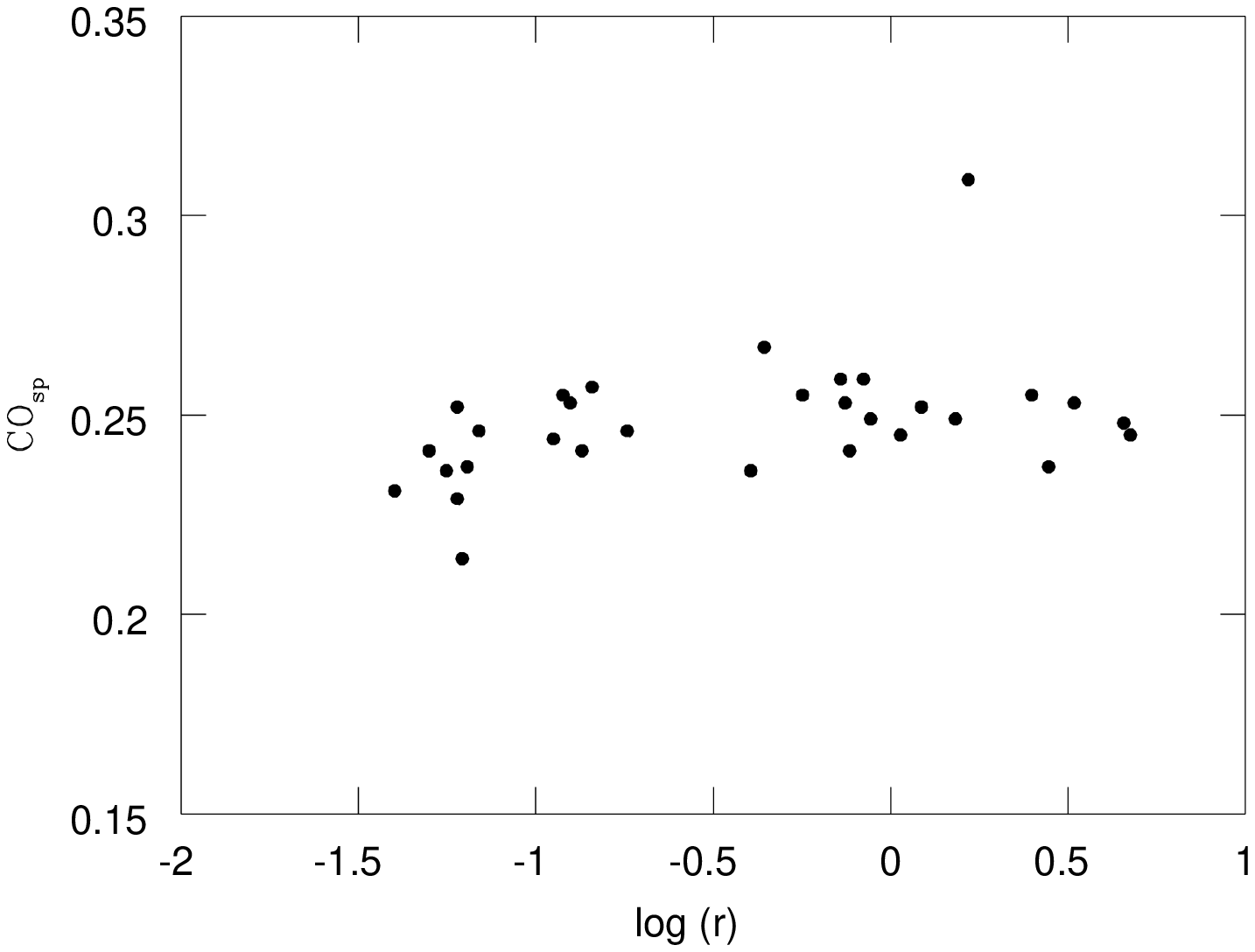,width=0.9\textwidth,angle=0}}
\caption{Changes of CO$_{sp}$ with radial distance to ellipticals 
 from the
center of the Coma cluster.}
\end{figure}

To further study the environmental dependence of the stellar population in
elliptical galaxies, the $< CO_{sp} >$ values of 30 ellipticals in the
Pisces, A2199 and A2634 clusters at $z=0.07$ (James and Mobasher 1999), are
compared to those in the Coma in figure 3c and Table 2. 
The $< CO_{sp} >$ values for ellipticals in $z=0.07$ clusters agree well 
with those at the core of the Coma cluster, indicating that they mostly consist
of old population, unlike the galaxies in less dense environments, 
although the statistical significant of this result is rather marginal. 

Since the present sample consists of elliptical galaxies (ie. no
lenticulars), the observed trend is not likely to be due to a dramatic
change in the galaxy population (eg. transition from SOs to
ellipticals) with radial distance from the center of the cluster.
However, it is more probable that the ellipticals in the outer parts
of the Coma cluster are undergoing final stages of evolution before
they stop star formation activity, while falling into the core of the
cluster. Indeed, recent observations at the peripheries of
high-redshift clusters ($0.2 < z < 1$), show a remarkable radial
gradient in the distribution of colour, Balmer absorption
lines and equivalent widths (Abraham et al 1996). Particularly
striking is the gradient in the $H\delta$ strong systems,
usually classed as ``post-starburst'' galaxies whose strong Balmer
lines result from a sharp truncation in their rate of star formation.
Moreover, observations of distant clusters have shown transitions of
SO population to ellipticals towards the core of the clusters
(Dressler et al 1998). Also, recent spectroscopic study of early-type
galaxies in the nearby Fornax cluster has revealed that the SOs have a
relatively younger age than the ellipticals (Kuntschner \& Davies
1998). Therefore, it is possible that the ellipticals, observed in the
periphery of the Coma cluster, are local counterparts of the
``post-starburst'' (E+A) population or SO galaxies (undergoing latest
stages of their evolution) observed at higher redshifts.

\section {Sources of scatter in the 
near-infrared fundamental plane of ellipticals}

By extending the FP of elliptical galaxies to near-IR wavelengths, it is
expected to minimise contributions from the young population and metallicity
in this relation and hence, reduce the 
observed scatter (Pahre et al 1998; Mobasher et al 1999).  
However, it has been discovered that the observed scatter
is not significantly reduced in the near-IR FP compared to 
its optical counterpart. A likely explanation is the relative 
contributions from the intermediate age AGB and red supergiant stars
to the near-IR light of ellipticals (Mobasher et al 1999). This is 
explored in the present section, using the $CO_{sp}$ strengths to measure
the contribution from these stars. 

Using the sample of 31 elliptical galaxies in the Coma cluster
(Mobasher et al 1999), for which
CO$_{sp}$ measurements are available from the present study, 
a three parameter plane fit was carried out to the K-band effective
surface brightness ($<SB_K>_e$), effective diameter ($A_e$) and
velocity dispersion ($\sigma$). This gives

$$log\ A_e = 1.48 log (\sigma) + 0.30 < SB_K>_e - 7.24 $$
with an {\it rms} scatter (in $log\ A_e$) of 0.077 dex. An edge-on view
of near-IR FP is shown in Figure 5. 

\begin{figure} 
\centerline{\psfig{figure=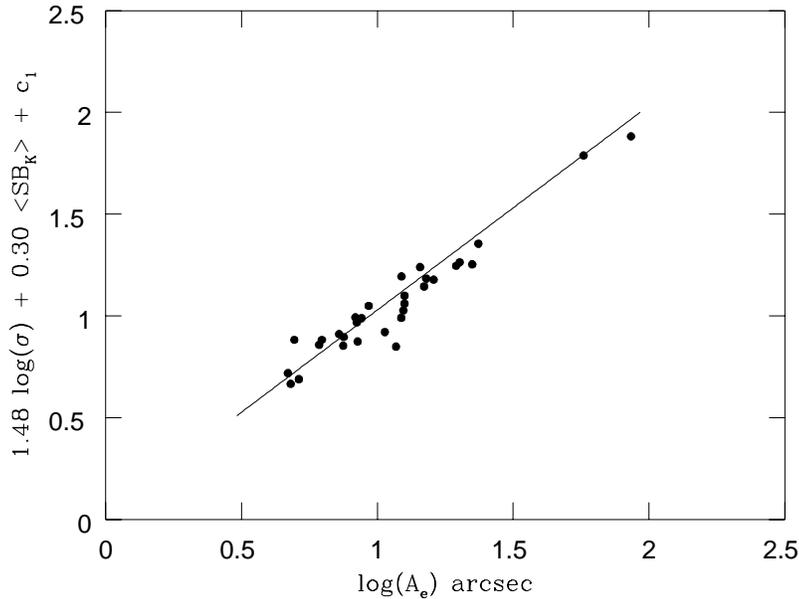,width=0.9\textwidth,angle=0}}
\caption{Near-IR FP of elliptical galaxies in the Coma cluster, using 31
galaxies listed in Table 1. The photometric data are taken from
Mobasher et al (1999). The line is a three parameter plane fit
to the data.}
\end{figure}

The residuals  from the the mean FP
at a given $log(A_e)-$
($\Delta (FP) = 1.48 log (\sigma) + 0.30 < SB_K>_e - 7.24 -log\ A_e $), 
are estimated for individual galaxies and 
are found to be correlated with their $CO_{sp}$ (Figure 6a). A least squares
fit to this relation gives $\Delta (FP) = -0.858\ CO_{sp} + 0.242 $ with
$98\%$ likelihood of this being a true relation. This trend is also confirmed
using CO$_{EW}$ values for the Coma ellipticals from Table 1 (Figure 6b). 
However, no significant correlation is found between 
$\Delta (FP)$ and $Mg_2$ strengths (Figure 6c). 
The presence of a relation between $\Delta(FP)$ and CO$_{sp}$ implies that
the observed scatter on the near-IR FP of Coma ellipticals (Figure 5) is, 
at least partly, due to changes in the contribution from the giant and
supergiant populations and is not a metallicity effect, as revealed
from the absence of a relation between $\Delta(FP)$ and Mg$_2$. 
This is consistent with the results from the previous
section in which ellipticals in the Coma cluster were found to have
different stellar populations, depending on their position 
(local density) in the cluster. Using the $\Delta (FP)\ vs.\ CO_{sp}$
relation (Figure 6a), 
the near-IR surface brightness of the galaxies in the present
sample are corrected for contributions from the giant and supergiant 
populations. 
This reduces the {\it rms} scatter on the near-IR FP from 0.077 dex 
(Figure 5) to 0.073 dex. 

\begin{figure}
\centerline{\psfig{figure=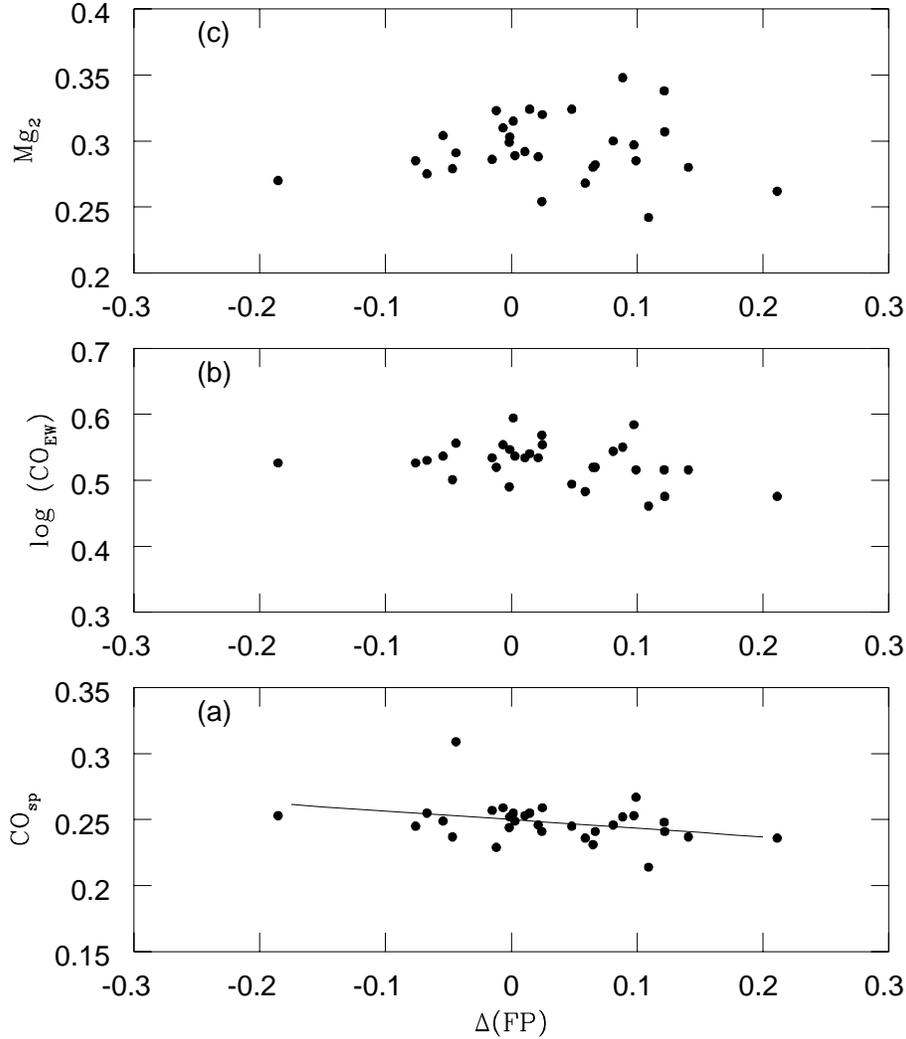,width=1\textwidth,angle=0}}
\caption{(a). Relation between the residuals from the near-IR FP
($\Delta(FP)$) and CO$_{sp}$ for the Coma ellipticals; (b). the same
as in Figure 6a for CO$_{EW}$; (c). the same as Figure 6a for Mg$_2$. 
}
\end{figure}

Therefore, although the giant and supergiant stars make a measurable 
contribution to the scatter on the near-IR FP of ellipticals,
their effect is rather marginal. The conclusion is that most of
the observed scatter in the near-IR FP of ellipticals is either caused by
observational errors or, is due to
dynamical effects and non-homology among the ellipticals (Pahre et al 1998). 

\section {The Relation between CO$_{sp}$ and Optical-IR 
Colours of Ellipticals } 

The $CO_{sp} vs. V-K $ relation for 31 ellipticals in the Coma cluster
(Table 1) is presented in Figure 7 with the coefficients of its
best linear fit listed in Table 3. The slope of $0.036\pm 0.007$ found here, 
is significantly shallower than 
$0.167\pm 0.023$, estimated for the Mg$_2-(V-K)$ relation
(derived from 47 ellipticals with available such data from
Mobasher et al 1999), which represents changes 
in metallicity and relative contributions from young and old
stellar populations among the ellipticals.  

The metallicity sequence on the CO$_{sp}\ vs. \ (V-K)$ diagram,  
as predicted from stellar synthesis models (Worthey 1994),   
is also shown on Figure 7. 
To calculate the CO$_{sp}$ 
features, not given in these models, the strength of Mg$_2$ lines
at any given $V-K$ colours are taken from Worthey (1994) models and
converted to CO$_{sp}$, using the empirical Mg$_2-$CO$_{sp}$ relation  
presented in section 3. However, one should note that the main source of 
uncertainty here is the lack of 
proper modelling of the AGB and supergiant populations in these models.   

\begin{figure} 
\centerline{\psfig{figure=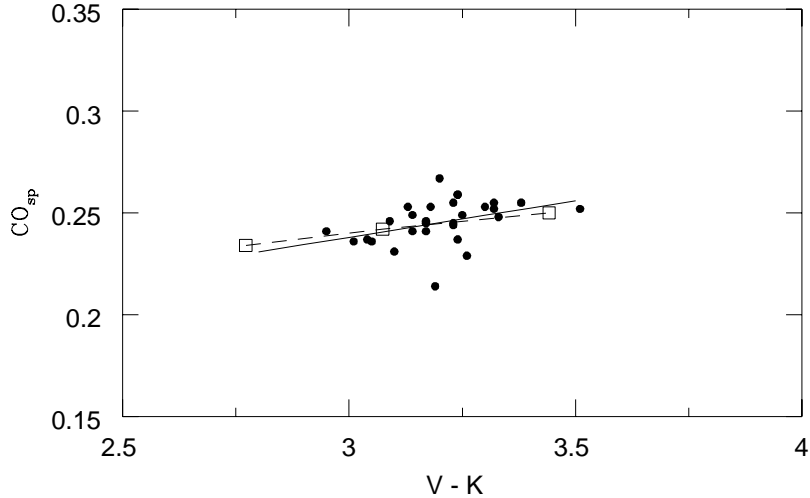,width=0.9\textwidth,angle=0}}
\caption{The CO$_{sp}-(V-K)$ relation for elliptical galaxies in the Coma
cluster. Solid line is the least squares fit to the data. 
Predictions from stellar synthesis models of Worthey (1994), corresponding
to a change in metallicity in the range $-0.5 < [Fe/H] < 0$ (increasing
metallicity towards redder $V-K$ colours)- (empty boxes) is indicated. 
An age sequence also follows the same trend. The model is
normalised to the data so that $[Fe/H]=0 $ corresponds to $V-K=3.44$
and CO$_{sp}=0.25$.  
}
\end{figure} 
 
The slope due to the metallicity sequence (in the range $-0.5 < [Fe/H] < 0$) 
in Figure 7 (dashed line), as predicted by stellar synthesis models, 
is close to the observed  
CO$_{sp}\ vs. \ (V-K)$ relation (solid line in Figure 7). This is also 
similar to an age trend, as predicted from the same models, indicating the
age/metallicity degeneracy.  

Using the present data, it is not possible to determine whether
age or metallicity dominate the relation shown in Figure 7, with the
models showing that both are capable of producing effects of the
observed size. However, the combined effect of age and metallicity appears
to be significantly reduced on the CO$_{sp}-(V-K)$ diagram, as compared
to that for Mg$_2-(V-K)$ relation (Table 3).  

\begin{table}
\caption { Coefficients for a linear least squares fit 
to the CO$_{sp}-(V-K)$ relation; $Y = a * X + b$}
\begin{tabular}{lcccc}\\ 
 & & & &   \\
             
  Y        & $X$    & $  a$ &  $  b$ & $   n$\\
       &        &                   &                 &   \\
 CO$_{sp}$ & $ V-K$ &  $0.036\pm 0.016$ & $0.130\pm 0.03$ & 29\\
       &        &                   &                 &   \\   
 Mg$_2$  &  $ V-K$  &  $0.167\pm 0.023$ &   $-0.240\pm 0.061$ &  47\\
       &        &                   &                 &   \\   

\end{tabular}
\end{table}

\section{Conclusions}

The  CO$_{sp}$ (2.3 $\mu$m) 
absorption features are estimated from the near-IR spectra of
a sample of 31 elliptical galaxies at the core 
and outskirts of the Coma cluster. Combined with other spectroscopic
($\sigma$ and Mg$_2$) and photometric (
K-band) data for this
sample, a study of the stellar population in elliptical galaxies
is carried out. The main conclusions from this study are summarised as follows:
\begin{enumerate}  
\item The mean CO$_{sp}$ values for elliptical
galaxies at the core of the Coma cluster are found to be smaller compared
to their counterparts in the outer region. There is a 
probability of $< 1\%$ for these galaxies belonging to the same parent
population. This is interpreted as due to the presence of
intermediate-age red giant and supergiant stars in ellipticals in low
density environments. This implies that the ellipticals in the outskirts
of rich clusters are relatively younger than their counterparts at the core.

\item Using the  CO$_{sp}$ values, the near-IR FP of ellipticals is corrected
for contributions from the intermediate-age supergiant stars. This reduces
the {\it rms} scatter in this relation (at a given
$log(A_e)$) from 0.077 dex to 0.073 dex. 
This modest reduction means 
that the observed scatter
on the near-IR FP of ellipticals is mostly dominated by 
observational errors or dynamical effects
and non-homology of ellipticals. 

\item A relation is found between CO$_{sp}$ and $V-K$ colours 
of ellipticals in 
the Coma cluster. The slope of this relation ($0.036\pm 0.016$) is
significantly shallower than that for the Mg$_2-(V-K)$ relation. 
Comparing with the stellar population models, it is found that
the observed trend is caused by   
either metallicity or stellar population changes.   

\end{enumerate}

\end{document}